\documentclass[preprint]{aastex}

\def\kms{\ifmmode {\rm \ km \ s^{-1}}\else \rm km \ s^{-1}\fi}
\def\cm{\ifmmode {\rm \ cm }\else $\rm cm$\fi}

\received{21 November 2000}
\accepted{.. .. 2000}

\begin{document}

\title{Ly$\alpha$ line formation in starbursting galaxies \\
I. Moderately thick, dustless, and static \ion{H}{1} media}

\author{Sang-Hyeon Ahn$^1$, Hee-Won Lee$^2$ and Hyung Mok Lee\altaffilmark{1,3}}
\affil{$^1$School of Earth and Environmental Sciences, Astronomy Program, Seoul 
National University, Seoul 151-742, Korea.}

\affil{$^2$Department of Earth Science, Sejong University, Seoul 143-747, Korea.}
\email{sha@astro.snu.ac.kr}

\altaffiltext{3}{Also at the Institute of Space and Astronautical Science, Japan}

\begin{abstract}
We investigate the Ly$\alpha$ line transfer
in nearby and high redshift starbursting galaxies,
where the effects of high optical depths and the role of dust
in the scattering medium are expected to be conspicuous and should be treated
in a very careful manner. We present our first results
in a dustless, static, and uniform \ion{H}{1} media with
moderate Ly$\alpha$ line center optical depths $\tau_0=10^{3-6}$.
We assume the temperatures of the media to be $T=10^{1-4}{\rm\ K}$,
and use a Monte Carlo technique. We investigate the basic processes
of the line transfer and confirm the criterion of
$a\tau_0>10^3$ for the validity of the diffusion approximation
suggested by Neufeld in 1990, where $a$ is the Voigt parameter.
Adopting the model for evolution of a galactic supershell suggested
by Tenorio-Tagle et al., we performed calculations
on the Ly$\alpha$ line formation for each evolutionary stage
of an expanding supershell.
The emergent Ly$\alpha$ profiles are characterized by the double peaks and
the absorption trough at the line center.
We found that the absorption troughs expected in most of the 
evolutionary stages are not wide enough to be observed with 
current instruments. However, the absorption trough in 
the Ly$\alpha$ emission profile
from an expanding recombining supershell can be marginally detected.
\end{abstract}

\section{Introduction}
With the advent of large ground-based and space telescopes,
primeval galaxies at high redshifts ($3<z<5$) have been detected through
a variety of methods. One of the most efficient detection
methods is the Lyman break method which can be used to select high redshifted
objects. Subsequent 
spectroscopy revealed the existence of
Ly$\alpha$ features in either emission or absorption
\citep{ste96,ste99} in many of galaxies showing Lyman break.

As well as being a redshift indicator, Ly$\alpha$ bears information 
on the physical state of those galaxies.
The emergent Ly$\alpha$ profile is very sensitive to  various factors
including the kinematics, scattering optical depth, and dust content.
Hence, it is important to investigate the Ly$\alpha$ line formation
for its cosmological applications.

The Ly$\alpha$ profiles of many Lyman break galaxies
in the early Universe may be classified into three types :
(1) symmetric profile,
(2) asymmetric or P Cyg type profile,
(3) broad absorption profile extending to the Lorentzian wings.
It is known that the Ly$\alpha$ profiles of nearby starbursting galaxies
also fall into one of the above three types \citep{kun98}.
It seems to be a general consensus that the P Cyg type Ly$\alpha$
profile is attributed to the expanding neutral medium surrounding
a Ly$\alpha$ source \citep{lee98,kun98,ahn00}.
This is supported by the existence of the
interstellar absorption from low ionization species such as
\ion{O}{1} $\lambda$1302 and \ion{C}{2} $\lambda$1335 which are
blueshifted relative to the Ly$\alpha$ line center \citep{kun98,spi99}.
It is also probable that the profile is formed in galactic superwinds.
Assuming that a galactic superwind possesses a Hubble
type outflow, \citet{lee98} calculated the emergent Ly$\alpha$ 
profiles and investigated their
spectropolarimetric properties.

The P Cyg type Ly$\alpha$ profile can also originate from a supershell
which is a remnant of strong explosions in a central star-forming cluster.
From these pieces of observational evidence, \citet{ten99}
proposed a model for the evolution of galactic supershells
based upon hydrodynamical simulations and ionization calculations.
However, they presented only a schematic description of the Ly$\alpha$
profiles in each evolutionary stage, so more systematic and quantitative
investigations are desired.

Ly$\alpha$ is one of the strongest resonance lines found in various
emission line objects and therefore has been the subject of intensive
research for a long time. Both numerical and analytical studies
have been carried out by a number of authors.
Recent advances in computer technology
have made it possible to explore the parameter space
that was not previously accessible.

The main objective of this study is therefore to understand
the Ly$\alpha$ line formation in a thick
medium by using a sophisticated Monte Carlo method, and to apply
the results to the models for the evolution of superbubbles
in starbursting galaxies.
This is the first paper in a series in which the
Ly$\alpha$ line transfer in an optically thick medium is investigated.
In this paper, we consider only moderately thick and dustless media,
and defer the cases of extremely thick media with and without dust
to subsequent papers.

This paper is composed as follows.
In Section 2, we describe the model for starbursting galaxies 
and our Monte Carlo code.
In Section 3, we show our results, and in the following section
we discuss the astronomical applications of our works. The final section
summarizes our major findings.

\section{Model and Method}
\subsection{Model}

Throughout this paper we consider a uniform, static, and plane parallel
medium consisting of pure neutral hydrogen atoms.
We characterize the thickness of the medium by the
Ly$\alpha$ line center
optical depth denoted by $\tau_0$ from the midplane to either side
in the $z$-direction.
We assume that a young and massive star cluster is located at the center
of the slab. This model describes the young starbursting
galaxies detected by the Lyman break method.

Recently \citet{ten99} modeled the starbursting galaxies by
a young star-forming cluster surrounded by a supershell, which
evolves with age. According to their Fig.~8, we can summarize
the evolution of superbubbles in starbursting galaxies as follows.

\noindent
Stage (a) - In the initial stage of starbursting galaxies, the surrounding
medium is neutral and static with respect to the central photon source.
During this phase ($t_{age}<1.5 {\rm\ Myrs}$), the ultraviolet photons
from the central cluster ionize the surrounding neutral medium.
The supernova explosions also deposit the kinetic energy
to the medium surrounding the star cluster. Subsequently an expanding shell,
which captures the UV photons inside the shell, develops.\\
Stage (b) - The shell breaks up by the Rayleigh-Taylor instability,
and the UV photons can freely ionize the surrounding medium.
Its ionization front usually gets to the outer edge 
in $t_{age}\simeq 2.5 {\rm\ Myrs}$.
At this time, a conical H II region of partially ionized gas is formed.
The ionized medium starts to expand
due to the increase of the internal pressure and the mechanical energy
injected by the central star cluster, thereby forming a supershell.\\
Stage (c) - The supershell becomes radiative and emits Ly$\alpha$ photons.
The neutral column density of the expanding
shell increases and its expansion speed decreases.\\
Stage (d) - Finally the neutral supershell forms, and the expansion
almost ceases.

In view of the above scenario, the starbursting
galaxies are composed of the central Ly$\alpha$ emitter or a young
star-forming cluster and the surrounding media.
There are three ingredients in modeling starbursting galaxies.
The first is the hydrodynamical evolution of a supershell,
the second is the photoionization, and the last is
the state of the photon source.
The former two ingredients are reflected in \citet{ten99},
and the last one opens two possibilities in the input spectrum.
When the major contributers of UV photons (i.e., O, B stars) are still alive,
the input spectrum consists of the Ly$\alpha$ emission and the UV continuum.
Therefore, in order to model the Ly$\alpha$ line formation in
each evolutionary stage of the starbursting galaxies,
we consider following six types of configurations.

\noindent
Case I - the scattering medium is neutral and
cold ($T=10 -100 {\rm\ K}$), and the Ly$\alpha$ source is located
at the center of the slab.\\
Case II - the scattering medium is partially ionized and hot ($T=10^4{\rm\ K}$),
and the Ly$\alpha$ source is located at the center of the slab.\\
Case III - the scattering medium is partially ionized and also
acts as a Ly$\alpha$ emission source. \\
Case IV - a continuum source is located at the center of
the scattering medium which is partially ionized and has
a temperature $T=10^4\ {\rm K}$.\\
Case V - a continuum source is located at the center
of the scattering medium which is cold with temperature $T=10\ {\rm K}$.\\
Case VI - there is a Ly$\alpha$ source at the center and
the scattering medium is an expanding neutral shell with $T=10\ {\rm K}$.

Stage (a) is described by Case I and Case V, because
the source of Ly$\alpha$ and UV continuum
is surrounded by the static, cold ($T=10{\rm\ K}$), and extremely thick
($\tau_0=10^{6-9}$) neutral medium.
Unfortunately, our code becomes very slow when $a\tau_0 > 10^3$,
where $a$ is the Voigt parameter.
We need a special scheme to overcome this limitation. In the present
study, we consider cases with $\tau_0 = 10^{3-6}$ only.

Assuming that the Ly$\alpha$ photons in Stage (b) are generated in the
central star-forming region that is partially ionized, this can be modeled
by Case II and Case IV. 

Stage (c) has two Ly$\alpha$ emission sources.
The major one is the central star cluster, and the minor one is the
expanding radiative supershell. The major emission source is similar to that
in Stage (b), and thus can be modeled by Case II and Case IV before
the \ion{H}{1} column density of the supershell becomes large
enough to produce the P Cyg type absorption blueward of the major peak.
The conical \ion{H}{2} region of Stage (b) can be described by Case III.
Case III also can describe the line formation in \ion{H}{2} regions where
young stars and interstellar media are all mixed.

In the earlier epoch of Stage (d), the expansion speed of the neutral
supershell is large, and there appears a P Cyg type Ly$\alpha$
emission line. This era can be described by Case VI.
Case V models the latter epoch of Stage (d) when the Ly$\alpha$ sources or
O stars have been extinguished. However, extremely thick cases will be
briefly treated in the current work and more details will be presented
in the future (see also \citet{lee98} for the galactic superwind).

\subsection{Previous Studies of Ly$\alpha$ Transfer}

Since we are interested in the emergent Ly$\alpha$ profiles in various
situations, we first introduce the dimensionless parameter $x$ defined by
\begin{equation}
x \equiv \Delta\nu/\Delta\nu_D=(\nu-\nu_0)/\Delta\nu_D,
\end{equation}
which describes the frequency shift from the line center $\nu_0$ in units
of the Doppler shift $\Delta\nu_D\equiv \nu_0 (v_{th}/c)$.
Here $v_{th}$ is the thermal speed of the scattering medium and
$c$ is the speed of light.

The Ly$\alpha$ optical depth for a given system is
\begin{eqnarray}
\tau_x = \tau_0 H(x,a),
\end{eqnarray}
where $H(a,x)$ is the Voigt function or Hjerting function,
$a=4.71\times10^{-4}T_{4}^{-1/2}$ is the Voigt parameter,
and $T_{4}$ represents the temperature of medium in units of $10^4{\rm\ K}$.
The line center optical depth is related to the \ion{H}{1} column density
$N_{HI}$ via
\begin{equation}
\tau_0 \equiv 1.41\ T_{4}^{-1/2}
\left[{N_{HI} \over {10^{13} \rm cm^{-2}}}\right].
\end{equation}
Since the normalized profile function $\phi (x)$ is related to the
Voigt function by
\begin{eqnarray}
\phi(x) = {1 \over \sqrt{\pi}} H(x,a),
\end{eqnarray}
the line center optical depth used by previous authors ($\tilde{\tau_0}$)
is related with that defined in this paper ($\tau_0$) by
$\tilde{\tau_0} = \sqrt{\pi} \tau_0$.

Theoretical studies on the Ly$\alpha$ line transfer
in an optically thick and static medium have a long history.
\citet{unn55} formulated the Ly$\alpha$ line transfer problem,
and \citet{ost62} proposed a simple physical picture
for understanding the resonance line transfer in a thick medium.
\citet{ada72} revised Osterbrock's picture 
and gave a detailed explanation of the transfer mechanism. Noting
that each scattering event is accompanied by an average shift in frequency
of order the thermal Doppler width $\Delta\nu_D$, he emphasized
the diffusive nature of the transfer process both in frequency space
and in real space introducing the terms `single longest flight' and
`single longest excursion.'

\citet{neu90} divided the optical depths into three regimes:
the slightly thick optical depth $\tilde\tau_0\le 10^3$,
the moderately thick optical depth $10^3<\tilde{\tau_0}<10^3/a$,
and the extremely thick optical depth $\tilde\tau_0\ge10^3/a$.
Here the range of $a$ is between $10^{-5}$ and $10^{-2}$
for a scattering medium with $10 < T < 10^4{\rm\ K}$.

Adopting the diffusion approximation, an analytic solution for the extremely
thick case was given by \citet{har73}, and a more general solution was
obtained by \citet{neu90}. In moderately thick media, however,
it is difficult to solve the problem analytically,
because the diffusion approximation is not valid in this regime.
Hence, a Monte Carlo method is useful to solve the problem of
the Ly$\alpha$ transfer in moderately thick media.

\citet{ave68} investigated the resonance-line scattering
by a Monte Carlo method. In their study, they focused on the difference
between the complete redistribution and the partial redistribution.
Their calculations reach the condition $\tilde{\tau_0} =10^7$
and $a=4.3\times10^{-4}$, which turned out to be about the critical
optical depth between the extremely thick and the moderately thick cases.
\citet{gou96} also investigated a similar radiative
transfer problem. However, they concentrated on the observational
feasibility including extinction in a partially ionized envelope
of Lyman limit galaxies, and the detailed line formation mechanism
was not fully described.
In addition to these works, there have been several investigations
using a Monte Carlo method for the resonance line transfer
in an optically thick and static medium
in the literature (e.g. Meier \& Lee 1981).

\subsection{The Monte Carlo Technique}

In this study, we investigate the problem by means of
a Monte Carlo technique. A detailed description of this code
was presented in the previous paper \citep{all00}.
Compared with those of other previous investigators,
the major advantage of our code is that it incorporates
all the quantum mechanics associated with both resonant and
wing scatterings of Ly$\alpha$.
Here, we neglect any collisional broadening in the upper state,
and assume that the frequency of the scattered photon is identical
to that of the incident photon in the rest frame of the scattering
atom.  This is Case II frequency redistribution in the atom's frame
introduced by \citet{hum62} (see also Mihalas 1978).
Hence, in the code we transform the necessary quantities
between the observer's rest frame and the scattering atom's rest frame.

According to \citet{ste80}, the scattering phase function associated
with wing scattering of Ly$\alpha$ is identical with the classical Rayleigh
scattering which yields 100 \% polarization for $90^\circ$ scattering.
However, the resonant scattering of Ly$\alpha$
associated with the transition $1s_{1/2}-2p_{1/2}$ is characterized by
the isotropic scattering phase function yielding totally unpolarized
scattered radiation, whereas for the $1s_{1/2}-2p_{3/2}$ transition
the scattering phase function is characterized by
the maximum polarization $p_{max}=3/7$ for $90^\circ$ scattering of
unpolarized incident photons. Therefore, in our Monte Carlo code,
we computed the probabilities for these scattering types
at each scattering event.

The absorption profile can be written as the oscillator
strength-weighted sum of two Voigt functions, 
\begin{equation}
\phi_{abs}(\nu)={1\over3}{1\over \Delta\nu_D\sqrt{\pi}}H(a,u_1)+
{2\over3}{1\over \Delta\nu_D\sqrt{\pi}}H(a,u_2),
\end{equation}
where
\begin{eqnarray}
u_1 &\equiv& {\nu-\nu_1 \over \Delta\nu_D} \nonumber \\
u_2 &\equiv& {\nu-\nu_2 \over \Delta\nu_D},
\end{eqnarray}
and $\nu_1$ and $\nu_2$ are the resonance frequencies corresponding
to the $1s_{1/2}-2p_{1/2}$ and $1s_{1/2}-2p_{3/2}$ transitions.

The scattering type can be determined from the scatterer's local
velocity, of which the distribution $f(y)$ is given by
up to normalization
\begin{equation}
f(y) = N_v {a\over 3\pi}\left({e^{-y^2}\over a^2+(u_1-y)^2}
+{2\ e^{-y^2}\over a^2+(u_2-y)^2}\right).
\end{equation}
Here $y\equiv v/v_{th}$ is the velocity in units of the thermal velocity, and
$N_v$ is the normalization constant. From the definition of the Voigt
function, we see that
\begin{equation}
N_v=3[H(a,u_1)+2H(a,u_2)]^{-1}.
\end{equation}

We may loosely say that the scattering is resonant with the transition
$1s_{1/2}\rightarrow 2p_{1/2}$ when in the rest frame of the scatterer
the incident photon has the wavelength within several natural widths,
say $10a$; i.e., $|u_1-y|< 10a$. Therefore, the probability
for a $1s_{1/2}\rightarrow 2p_{1/2}$ resonance transition is given by
\begin{eqnarray}
P_{r1} &\simeq& \int_{u_1-10a}^{u_1+10a}\ dy\ f(y) \nonumber\\
&\simeq& e^{-u_1^2}/[H(a,u_1)+2H(a,u_2)],
\end{eqnarray}
because the change in the exponential part is negligible in the
integration interval. A similar consideration gives
the probability for $1s_{1/2}\rightarrow 2p_{3/2}$ transition,
\begin{eqnarray}
P_{r2} &\simeq& \int_{u_2-10a}^{u_2+10a}\ dy\ f(y) \nonumber \\
&\simeq& 2e^{-u_2^2}/[H(a,u_1)+2H(a,u_2)].
\end{eqnarray}
Therefore, the probability that a given photon is scattered
in the damping wings is
\begin{equation}
P_{nr}= 1-P_{r1}-P_{r2}.
\end{equation}

In the code we determine the scattering type in accordance with
the probabilities $P_{r1}, P_{r2}$ and $P_{nr}$.
If scattering is chosen to be resonant, then we set $u=x$.
Otherwise, the scattering occurs in the damping wings, and $u$ is chosen
in accordance with the velocity probability distribution given by Eq. (7).

We assign the propagation direction $\vec k_f$ of a scattered photon
in accordance with one of the three phase functions.
The scattered velocity component $v_\perp$
perpendicular to the initial direction $\vec k_i$
on the plane spanned by $\vec k_i$ and $\vec k_f$ is also governed by
the Maxwell-Boltzmann velocity distribution, which is numerically obtained
using the subroutine ${\it gasdev}()$ suggested by \citet{pre89}.
The contribution $\Delta x$ of the perpendicular velocity component
$v_\perp$ to the
frequency shift along the direction of $\vec k_f$ is obviously
\begin{eqnarray}
\Delta x = {v_\perp\over c} [1-(\vec k_i \cdot \vec k_f )^2]^{1/2}.
\end{eqnarray}
Therefore, the frequency shift $x_f$ of the scattered photon is given by
\begin{eqnarray}
x_f = x_i - u + u(\vec k_i \cdot {\vec k_f} )
    + {v_\perp\over c} [1-(\vec k_i \cdot \vec k_f )^2]^{1/2},
\end{eqnarray}
where $x_i$ is the frequency shift of the incident photon.

In each scattering event, the position of scattered photon is checked.
If the photon escapes from the medium or $|z|<\tau_0$, we collect 
that photon according to its frequency and escaping direction.
The frequency bin is chosen to be $0.25\le\Delta x\le2$ considering
the computing speed and the spectral resolution required to describe
the details of the emergent profiles.
The whole procedure is repeated to collect typically
about $10^3$ photons in each frequency bin.
This code, being very faithful to the quantum mechanics of
the scattering of Ly$\alpha$, we may compute the polarization
of Ly$\alpha$. This will be presented in the future (see also Lee \& Ahn
1998).

\section{Line Formation Mechanism}

\subsection{Emergent Profiles}

In this section, in order to test our Monte Carlo code, we compute emergent
profiles for the thick cases with various optical depths, and compare
them with the analytic solutions considered by \citet{har73}
and \citet{neu90}. They investigated the diffusive
process of the resonance line transfer in an plane-parallel,
extremely thick, and static medium with a 
Ly$\alpha$ source in the central plane.
We performed Monte Carlo calculations for the same cases,
and compared results with the analytic solutions. From the comparison,
we can determine the critical optical depth that divides the moderately
and the extremely thick optical depths,
where in the latter case the diffusion approximation is can be applied.

By adopting the Eddington approximation, \citet{har73} introduced
the diffusion equation for the angle-averaged intensity $J(\tilde\tau,x)$
\begin{equation}
{\partial^2 J\over\partial\tilde\tau^2}+{\partial^2 J\over\partial\sigma^2}
=-3\phi(x){E(\tilde\tau, x)\over 4\pi},
\end{equation}
where $E(\tilde\tau, x)$ is the photon generation rate (per unit mean optical
depth per unit Doppler width per unit area) and the frequency parameter
$\sigma$ is defined by
\begin{equation}
\sigma \equiv\sqrt{2/3}\int_0^x dx'/\phi(x').
\end{equation}

For the case of a monochromatic source at the line center
located at the midplane of the slab
with $E(\tilde\tau,x)=\sqrt{3/2}\delta(\tilde\tau)\delta(\sigma)$
and $(a\tilde{\tau_0})^{1/3} \gg1$,
\citet{har73,neu90} derived the following analytic solution
for the emergent mean intensity $J$,
\begin{eqnarray}
J(\pm\tilde{\tau_0},x) = {\sqrt{6} \over 24}{x^2 \over a\tilde{\tau_0}}
{ 1 \over \cosh[(\pi^4/54)^{1/2}(|x^3|/a\tilde{\tau_0})]}.
\end{eqnarray}

We performed Monte Carlo calculations for the same case, and
Fig.~1 shows our results. Notice that we omit the profiles in the
blue part because the profiles are symmetric with respect
to the origin ($x=0$). The solid lines represent the results
of our Monte Carlo calculations, and the dotted lines represent
the analytic results obtained by \citet{neu90}.
Our results for ($\tau_0=10^6$, $a=4.71\times 10^{-4}$) and
($\tau_0=10^4$, $a=1.49\times 10^{-2}$) are in good agreement
with the analytic solutions.
However, the result for the case ($\tau_0=10^5$, $a=4.71\times 10^{-4}$)
shows noticeable discrepancy. Our Monte Carlo solution appears to be
translated from the analytic solution toward larger $x$
by an amount of $\Delta x\simeq 1$, where wing scatterings become important
and $\phi(x)$ can be well approximated by

\begin{equation}
\phi(x) \simeq {a \over \pi x^2}.
\end{equation}

In the cases of dust free media, no loss of Ly$\alpha$ line photons is
permitted so that the flux is conserved.
The profile function $\phi(x)$ can not be approximated by Eq.~(17)
at $x\le 3$, but more adequately described by $\phi(x) \simeq
(a/\pi)/(1+x^2)$. Therefore, Neufeld's calculation underestimates
the number of core photons that are removed and ultimately
redistributed to the wing regimes.
This reduces the number of diffusively transferred wing photons,
and hence the emergent spectra show significant discrepancy
for moderately thick media.

According to \citet{neu90}, the diffusion approximation, Eq.~(17), is valid
only for $a\tilde{\tau_0} \ge 10^3$. For the former two cases,
we have $a\tilde{\tau_0}=562,\ 264$ for ($\tau_0=10^6,\ a=4.71\times 10^{-4}$)
and $(\tau_0=10^4,\ a=1.49\times 10^{-2}$), respectively. On the other hand,
$a\tilde\tau_0=56.2$ for $(\tau_0=10^5,\ a=4.71\times 10^{-4})$.
Therefore Neufeld's condition for the extremity of optical depth is marginally
valid in the former two cases, and this also confirms that our code operates
very well.

\subsection{Last Scattering Positions}

The last scattering positions of escaping photons are investigated in order
to understand the line transfer processes in detail. We denote by
$P_z$ the $z$-coordinate values of the last scattering
positions of escaping photons and plot $P_z/\tau_0$ in Fig.~2. 
We identify two kinds
of peaks in the distribution, one at the central region ($P_z/\tau_0\approx 0$),
and the other located near the boundaries ($|P_z|/\tau_0\approx 1$).
We also find that the number of photons escaping from the center decreases
as $\tau_0$ gets larger and the scattering medium colder.

According to \citet{ada72}, in a slightly thick medium (${\tilde\tau}_0<10^3$),
an initial core photon ($x_0=0$) experiences a large number of core
scatterings until being scattered at the wing frequency by an atom
that moves much faster than the thermal speed of the medium.
When a wing scattering happens, the photon is ready to escape
from the medium, and the escape is achieved by `a single longest flight'.
During the core scatterings, the mean free path of the photon is so small
that the photon can not travel much. Hence, these photons constitute
the central peak in Fig.~2.

On the other hand, in a moderately thick medium ($10^3<{\tilde\tau}_0<10^3/a$),
a wing photon can not escape from the medium by a single wing scattering,
but it can escape after only a few wing scatterings.
However, during these few successive wing scatterings, the photon may become
a core photon again by the so-called `restoring force,'
and then it experiences a large number of core scatterings.
While these processes are repeated, a photon wanders about the medium
before it escapes. By a Monte Carlo technique, \citet{all00} showed
that this process, which they named `a wandering,' indeed operates.
In this regime, the last scattering position of escaping photons
is distributed rather uniformly, as shown in Fig.~2.

In the case of extremely thick media, Ly$\alpha$ photons escape
by the transfer process similar to that for the moderately thick medium case.
They alternatively experience a large number of core
scatterings and a series of wing scatterings.
However, when the medium is extremely thick and the temperature
of the scatterers is low, wing scatterings become more important
in the line transfer and the number of successive wing scatterings
becomes large. Hence, the escape of photons is mostly achieved
during the series of wing scatterings. \citet{ada72} called this
process `excursions' and their escape is achieved
by `a single longest excursion'.

As a result, photons gradually smear out near the boundary of the medium,
and they escape by a single longest excursion.
These photons form the peak at $|P_z|/\tau_0\simeq 0.95$ and
the small portion in $0.2<|P_z|/\tau_0<0.8$ in Fig.~2.
In view of these excursion processes, we can understand
that the central peak is suppressed and the outer peaks dominate,
as $\tau_0$ of the medium gets larger and the temperature of
the scattering medium gets lower.

\subsection{Last Scattering Path Length}

Next, we investigate the distribution of photon path lengths
just before escape. In Fig.~3 we show our results,
where the horizontal axis represents the path length $\tau_f$
which is the optical depth traversed by an escaping photon in the slab
after its last scattering in units of $\tau_0$.
In Fig.~3 one may notice that there are three kinds of peaks.
The first peak appears at $\tau_f \approx 0$, the second at $1<\tau_f<2$, and
the third at $\tau_f \approx 2$.

We may argue that the first peak is formed by the photons
that have experienced `single longest excursions'.
Considering the processes mentioned in the previous subsection,
these photons have traveled in real space through excursions and
therefore escape happens preferentially for those photons that have diffused
spatially near the boundary of the scattering medium.

The second peak is due to the photons that escape
by the `single longest flight' from the central region of the medium.
We may explain the shape of the distribution curve for
this peak as follows. Since a large number of core scatterings isotropize
the radiation field, as we will show in Fig.~4,
the angular distribution of the emergent photons is approximately given by
$P(\mu) \propto \mu$ in the vicinity of the photon source.
Here, $\mu$ is the cosine of the angle between the outgoing wavevector and
the normal direction of the scattering medium.
Therefore, the escape by a single longest flight can happen in the direction
$\mu$ when the path length satisfies $\tau_f\simeq\tau_0/|\mu|$.
This implies that the path length distribution
$f(\tau_f=\tau_0/|\mu|)\propto |\mu|
\propto \tau_0/\tau_f$ for $\tau_f >\tau_0$,
where $\mu\in[0,1]$. In addition, a minor contribution to the second
peak is due to the photons last scattered
in the intermediate region, $0.2<|P_z|/\tau_0<0.8$.

The third peak is contributed by the back-scattered photons.
The frequency diffusion process drives
these back-scattered photons to acquire very large wing frequencies,
at which the medium becomes transparent enough for them to escape readily.
Hence, this peak appears only when the wing scatterings become important.
Therefore, as $\tau_0$ gets larger and the scattering
medium colder, the third peak gets stronger and the first and second peaks
become weaker. In particular, when $\tau_0=10^4$, $T=10\ {\rm K}$,
no second peak appears.

\subsection{Dependence of the Mean Scattering Number on the Initial
Frequencies}

In this subsection, we investigate the dependence of the mean number of
scatterings $\langle N \rangle$ on the initial frequency $x_0$ and
the Voigt parameter $a$. This was investigated by \citet{ave68}
using a weighting scheme. They introduced the weighting scheme
in order not to count those photons that are initially
in the wing regime and directly escape without being scattered, and hence
to prevent the $\langle N \rangle$ distribution from varying abruptly at 0.
However, we did not adopt this scheme, and show our results
in Fig.~5. Here we note that the areas under the curves are normalized
to the same value because no dust is present in the medium and
the weighting scheme is not used.

Although, at small $\langle N \rangle$, the emergent flux
appears rather fluctuating, our results
are in qualitative agreement with those obtained by \citet{ave68}.
We see that the number of directly escaping photons increases
as the initial photon frequency gets larger and the medium gets colder.

We explain these facts as follows.
The initial wing photons with $x_0\ge2.5$ experience coherent wing scatterings
and can escape more easily with a small $\langle N \rangle$.
However, the so-called `restoring force' renders some
of wing photons to be core photons, which will subsequently experience 
a large number of core scatterings. Therefore, we have two local maxima
in $\langle N \rangle$ of escaping photons
and we get less photons with large $\langle N \rangle$ when $x_0$ is large.
The larger fraction of directly escaping photons in the cold medium
means that the smaller width of the Doppler core
and the larger probability of wing scatterings in the Voigt profile
play a main role in the line transfer.

\section{Observational Ramifications}

In this section we apply our code to model
the Ly$\alpha$ line formation in starbursting galaxies.
We adopt the models described in Section 2, and show our results
of the Monte Carlo calculations.

\subsection{Case I - Hot Source and Cold Scatterers}

In Case I, a hot source is embedded in a cold scattering medium
with the line center optical depth $\tau_0=10^{1-4}$.
Here, two types of sources are considered. The first is the O, B stars
in a young star cluster, and the other is the central region ionized
by them. However, for both sources the input profile can be assumed to have
a width of about $10{\rm\ km\ s^{-1}}$. This is because a typical young
cluster has a one dimensional velocity dispersion
of $10<\sigma_v<20{\rm\ km\ s^{-1}}$,
and the typical temperature of the ionized region is $T=10^4 {\rm\ K}$
or $\sigma_v\approx 10{\rm\ km\ s^{-1}}$.
Due to the lack of our knowledge about the source, we just assume
a Gaussian profile
$\phi(x) \propto \exp(-x^2/x_s^2)$,
for the Ly$\alpha$ emission from the source, where $x_s\approx 30$. 
The value of $x_s$ was chosen such that the line width of source
is greater than the Doppler width of the scattering medium
($T=10{\rm\ K}$) by a factor of $10^{3/2}$. We have not considered 
the microturbulence.

We show our Monte Carlo results in Fig.~6.
The emergent profiles are characterized by double peaks and 
a central absorption trough, which are superposed on the broad source profile.
The full width of the absorption trough is found to be $\Delta x \approx 6$,
which corresponds
to $\Delta \lambda \simeq 0.01 {\rm\ \AA}$ in the rest frame.
The value of $\Delta\lambda$ is weakly dependent upon the temperature of
the scattering medium.

For those objects located at $z=3$,
the width of the absorption trough becomes $0.04{\rm\ \AA}$.
Considering that the spectral resolution of the Low Resolution Imaging
Spectroscope (LRIS) on Keck I is about $2-5 {\rm\ \AA}$
(Oke 1995), the detection of the absorption trough profile may
be impossible with LRIS, but it will be possible in the near future.

The total hydrogen column density of normal spiral galaxies and dwarf galaxies
ranges from $N_{HI}=10^{19-22}{\rm\ cm^{-2}}$, which corresponds
to the Ly$\alpha$ line center optical depth of $\tau_0=10^{6-9}$.
Hence for such realistic cases, a broader absorption trough is expected,
which can be more readily observable.
Moreover dust can play an important role in widening the absorption trough.
However, the Monte Carlo method becomes highly inefficient
for the extremely thick cases with $a\tau_0 > 10^3$, because the number of
core scatterings per excursion increase enormously. More detailed
investigation on this issue will be reported in forthcoming papers.

The two peaks around the central frequency are a consequence of the assumption
of the dustless medium. The photons constituting the peaks
have traversed very long distances.
In this case, they are prone to be destroyed by even a small amount of dust
that may co-exist in the scattering medium \citep{che94}.
Hence we may make use of their existence and strength to measure the degree
of metal contamination or the dust contents of the medium. We will report
the results on this topic in the future.

\subsection{Case II - Monochromatic Source and Hot Scatterers}

When the scattering medium is partially ionized and the central Ly$\alpha$
source is the O, B stars or the central dense \ion{H}{2} region,
we may safely assume that the source is monochromatic.
This is because the width of the source is similar to the Doppler width of
the scattering medium.
Hence, we put a monochromatic and isotropic source
with $x_0=0$ at the center of the plane-parallel medium,
where the temperature of the scattering medium is set
to be $T=10^4 {\rm\ K}$, typical value for \ion{H}{2} regions.

Our results are depicted in Fig.~7, where we show
the effects of various line center optical depths.
As $\tau_0$ gets larger, the peaks move farther away from the line center.
When $\tau_0<10^3$, the black absorption at the line center disappears
due to the slab geometry whose covering factor is unity.
We note that the moderate optical depth regime starts from $\tau_0>10^3$,
which can be seen from the appearance of the black troughs in Fig.~7.

We show the effect of the Voigt parameter on the line formation
in Fig.~8, where the case for $\tau_0=10^4$ is shown
in the lower panel, and the case for $\tau_0=10^3$ in the upper panel.
Here, since $a=4.71\times10^{-4}(T/10^4K)^{-1/2}$,
$a=4.71\times10^{-4}$ corresponds to $T=10^4 {\rm\ K}$,
$a=4.71\times10^{-3}$ to $T=10^2 {\rm\ K}$,
and $a=1.49\times10^{-2}$ to $T=10 {\rm\ K}$, which are representative
of the three phases of interstellar medium.

In the figure, we see that the peak gets broader and the peak center
moves farther from the line center as the scattering medium gets colder.
This is because the radiative transfer
is more affected by wing scatterings as $a$ gets larger.
We note that the increase of $a$ affects the line formation in a similar
way as the increase in $\tau_0$ in the extremely thick cases,
where wing scatterings dominate the line transfer.
We also note that the broad line spectra can be seen only
for examples where the parameters approximately satisfy the condition
$(a\tau_0)^{1/3}\gg1$, which is described in Section 3.1
and also discussed by \citet{neu90}.

In this case, $\Delta x = 1$ corresponds
to $\Delta \lambda \approx 0.05 {\rm\ \AA}$ in the emitter rest frame.
Therefore, the absorption trough has a width of $\Delta\lambda\approx 0.3{\rm\ \AA}$,
which is expected in the ultraviolet spectra of nearby starbursting galaxies.
When those objects are located at $z=3$,
the width of the absorption feature is about $1.2{\rm\ \AA}$ which can be
marginally detected by the Keck LRIS.

\citet{ten99}
predicted only the symmetric single peak in Stage (b).
We found from our calculations that Ly$\alpha$ may have double peaks
if the density of the central \ion{H}{2} region is high enough to
form a partially ionized medium.
However, the validity of this prediction is largely dependent upon
the existence of partially ionized zone in the star cluster.
This may be checked by exploiting hydrodynamical simulations and
calculating the ionization structure. We also note that dust may
seriously alter these features. However, these topics are beyond
the scope of this paper, and we concentrate on the radiative
transfer problem in a pure hydrogen medium.

\subsection{Case III - Uniformly Distributed Source}

We also performed calculations for the case of uniformly distributed
sources in the slab. This case can be applied to the recombining
shell at the epoch (c) in Fig.~8 of \citet{ten99}.
We distribute Ly$\alpha$ sources at $(P_x,P_y,P_z)=(0,0,\tau_s)$,
where $\tau_s=(2R-1)\tau_0$ and $R$ is a uniform random number in
the interval $[0,1]$. The number of sources are determined by the following two
constraints. The uniformity of the sources should be well described by
the Monte Carlo code, and the S/N ratio of the emergent profiles should
be high enough to give us meaningful results. In our Monte Carlo computations
approximately a few hundred photons per frequency bin with $\Delta x = 0.25$
are found to satisfy these constraints.

Each source radiates monochromatic Ly$\alpha$ photons isotropically.
Since the scattering medium is partially ionized, the temperature of
the medium is set to be $T=10^4{\rm\ K}$. Moreover, since the Lyman limit
opacity is smaller than the Ly$\alpha$ line center optical depth by a factor of
$10^{3-4}$, we performed the Monte Carlo calculations for
this case by setting the Ly$\alpha$ line center optical depth $\tau_0<10^5$.
In other words, we assume that this case can be applied
only before the recombination shell becomes extremely thick.

In Fig.~9 we show Ly$\alpha$ profiles for $10\le \tau_0 < 10^5$
and temperature $T= 10^4{\rm\ K}$.
In the figure we can see the effects of the line center optical
depth on the emergent profiles. When compared with the results
for the cases of the midplane source, no apparent
discrepancy was found for $\tau_0=10^4$.
For $10\le \tau_0 \le 10^3$, the inner part of the profile
gets broader, and the number of photons near the line center increases.
This is caused by the contribution of photons at the shallow region,
which can escape more easily without much frequency diffusion.
However, when the line center optical depth is large, frequency diffusion
takes place, and the emergent profiles become similar to the midplane cases.

In Fig.~10, we also show the effect of the Voigt parameter
on the line formation for the distributed source. We can see that
the emergent profiles for the distributed source have a stiffer inner
part of the peak than those for the midplane source. This trend is more
conspicuous in the case of $\tau_0=10^4$ than $\tau_0=10^3$.
In the case of $\tau_0=10^3$, the escape of photons takes place mostly
by single longest flights, and is less affected by the temperature
of the scattering medium.
However, when $\tau_0=10^4$ and $T<100{\rm\ K}$, wing scatterings
become dominant in the line transfer. Hence, photons smear out
near the surface of the medium, and escape easily without
much frequency diffusion. Therefore, the peak profile
shows a stiffer inner part and the maximum flux
occurs closer to the line center than in the midplane source case.

\subsection{Case IV \& V - Continuum Source}
When there are O, B stars alive in the central star-forming region,
their continua also contribute to the Ly$\alpha$ source photons,
in addition to the Ly$\alpha$ recombination line photons.
Thus far only the line source has been considered.
Now we consider the UV continuum as a photon
source. Then the emergent profile is the sum of these two components,
whose ratio is dependent upon the stellar evolution history of the
central cluster.

In Case IV, we assume that a continuum source with a flat spectrum
is located at the center of the slab, and that the scattering medium is
partially ionized or hot ($T=10^4{\rm\ K}$).
This case is meaningful only when the total line center optical depth
$\tau_0\le 10^4$, and therefore in this study we consider
$10 \le \tau_0 < 10^5$ in order to safely cover the meaningful range of $\tau_0$.

In Case V, we assume a flat UV continuum as the photon source and a
cold ($T=10{\rm\ K}$) medium as the scatterer.
In this case, the scattering medium is expected to have the line center
optical depth $\tau_0=10^{6-9}$, but due to the limitation in computing speed
of our present code, we perform calculations only for $\tau_0=10^{3-6}$.

The results for Case IV and V are depicted in Fig.~11 and Fig.~12, respectively.
For both cases IV and V, there appear double peaks with a broad absorption
trough at the line center. For Case IV, the width of the trough is
about $0.3-0.5{\rm\ \AA}$. When the objects are located at the redshift $z=3$,
the width of the absorption trough is about $1.2-2.0{\rm\ \AA}$, which
is marginally observable with the large ground-based telescopes.
The spectral features obtained in Case V are qualitatively similar
to those in Case IV, except for the smaller widths $0.01 - 0.03{\rm\ \AA}$.
When the source is located at the redshift $z=3$, the width
of the absorption trough ranges from $0.04-0.13{\rm\ \AA}$.
We expect that the absorption trough becomes much wider when we consider
the effect of dust absorption and the high \ion{H}{1} column density
which is more realistic than the models considered in this work.

\section{Summary and Discussion}

In order to predict the Ly$\alpha$ profiles from
starbursting galaxies, we investigated the Ly$\alpha$ line transfer
in a moderately thick and dustless medium using a Monte Carlo code.
Our code is faithful to the quantum mechanics associated with the
resonant and wing scattering of Ly$\alpha$, and the partial frequency 
redistribution is accurately treated.

We compared our results with the analytic solutions obtained by previous
investigators, and found that the results are in excellent agreement.
We confirmed the line transfer mechanism such as
a single longest flight, a wandering, and a single longest excursion
as proposed by \citet{ada72}.
Comparing our emergent profiles with the analytical solution of \citet{neu90},
we found that the wing approximation does not hold
for moderately thick media $10^3<\tilde{\tau_0}<10^3/a$,
and that it is valid only for the regime $a\tilde{\tau_0}>10^3$.

Next, we applied our results to the
Ly$\alpha$ line formation in starbursting galaxies,
in relation with the model given by \citet{ten99}.
The following is the summary of the emergent Ly$\alpha$ profiles
for each evolutionary stage of a superbubble in starbursting galaxies.
A brief summary is also
presented in Table~1, and the spectral capabilities of typical
instruments are shown in Table~2 for comparison.

(a) In the early phase of a starbursting galaxy,
the galaxy has a static and cold ($T=10\ {\rm K}$)
envelope, and the line width of the source can be set to be $10\kms$.
We obtained the emergent profiles consisting of double emission peaks
and an absorption trough
with width $\Delta\lambda\approx 0.01{\rm\ \AA}$ in the rest frame.
If this object is redshifted to $z=3$, the width is widened
to $\Delta\lambda\approx 0.04{\rm\ \AA}$ in the optical band,
which is beyond the observational limit of current instruments.

(b) The surrounding medium subsequently forms a conical
\ion{H}{2} region, and a partially ionized
envelope may be formed at the center.
We assume that the scattering medium is hot ($T=10^4\ {\rm K}$)
and the source is at the center of the slab.
The emergent Ly$\alpha$ profiles show symmetric double
peaks and an U-shaped absorption trough at the line center.

The trough has a width of $\Delta\lambda\approx 0.2{\rm\ \AA}$,
and the features that are formed in nearby starbursting galaxies
are found to be observable using the current high resolution far UV spectroscopy
such as HST/STIS. When those objects are located at $z=3$,
the width of the absorption feature is about $1{\rm\ \AA}$ which 
can be marginally detected with the Keck LRIS.

(c) In the next stage, an expanding superbubble develops.
In this stage, there appear two Ly$\alpha$ emission peaks.
The major peak is formed by the central star cluster and
the minor peak is formed by the radiative and expanding supershell.
Hence, the major peak is located at the line center,
and the minor peak is in the blue part of the line center.
According to our investigation, the major and minor peaks may
have an absorption trough at the center.

The width of the trough corresponds
to $\Delta \lambda \approx 0.25{\rm\ \AA}$.
Hence, this central U-shaped
absorption feature in nearby starbursting galaxies may be marginally
observable using the HST/STIS.
When those objects are located at $z=3$,
the width of the absorption feature becomes $\sim 1{\rm\ \AA}$,
and this trough should be marginally detected
by using ground based large telescopes.

(d) When the neutralization of the supershell is complete,
its \ion{H}{1} column density reaches $N_{HI}=10^{19-22}{\rm\ cm^{-2}}$,
and its expansion gradually ceases. Hence, we modeled
this stage by a hot source surrounded by a static and cold
($T=10\ {\rm K}$) scattering medium.
Since at this stage the central Ly$\alpha$ source is extinguished,
we considered the ultraviolet continuum source near Ly$\alpha$.
The profiles emergent at this stage are also characterized by
double peaks and U-shaped absorption trough.

In this paper, we concentrated on the cases of moderate optical depths.
However, the line center optical depth $\tau_0$ of the neutral envelopes
in starbursting galaxies is expected to be $10^{6-9}$.
In order for our model to be realistic, we must investigate these extremely
thick optical depths and include dust effects, which will enhance
the width of absorption trough, and make its detection more feasible.
We will study these cases in the future.

In section 3.3 we show the angular distribution $P(\mu)$ of escaping photons.
In the case of a slightly thick medium, we get an almost isotropic
distribution given by $P(\mu)\propto \mu$.
In a moderately thick medium, photons tend to escape more in the direction
parallel to the slab and hence $P(\mu)$ becomes convex upward.
In the extremely thick medium the escaping photons are emergent
preferentially in the direction normal to the plane, and $P(\mu)$ becomes
convex downward. This means that when wing scatterings dominate
the line transfer, the photons are affected by the geometrical shape of
the scattering region and the quantum mechanical properties of scattering.
Therefore we expect the development of polarization
when Ly$\alpha$ photons are radiatively transferred in an extremely
thick aspherical medium \citep{cha60,lee98}. This will be considered
in the near future.

\acknowledgments
HML and SHA are grateful for the financial support of Brain Korea 21
of the Korean Ministry of Education. HWL is also supported by
Brain Korea 21 awarded to Yonsei University. We thank Chris Pearson
for careful reading of the manuscript. This work was
supported by KOSEF Grant No. 1999-2-113-001-5.

\clearpage

\begin{figure}
\plotone{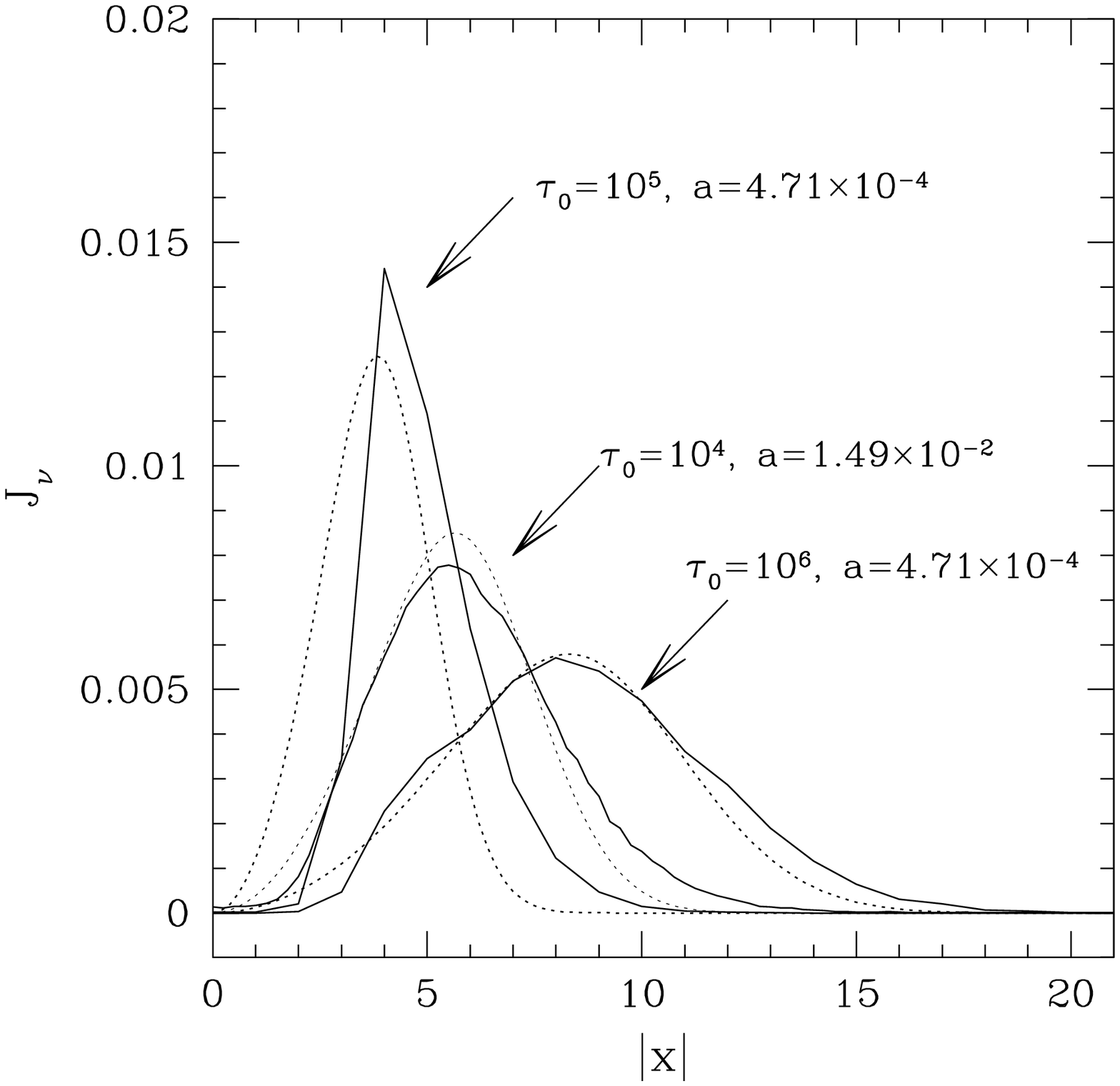}
\figcaption[f1.ps]{Our emergent Ly$\alpha$ profiles (solid lines) are
compared with Neufeld's analytic solutions (dotted lines).
The horizontal axis is the frequency in units of the thermal width,
and the total flux of the line is normalized to $1/4\pi$ in accordance with
Neufeld's normalization. The profiles are symmetric about the origin,
$x=0$. Here the Voigt parameter is related with the temperature
by $a=4.71\times10^{-4}T_{4}^{-1/2}$, where $T_{4}$ represents
the temperature of medium in units of $10^4{\rm\ K}$.
We note that the
our results are in good agreement with analytical results by Neufeld
for large $a\tau_0$.
\label{f1}}
\end{figure}

\begin{figure}
\plotone{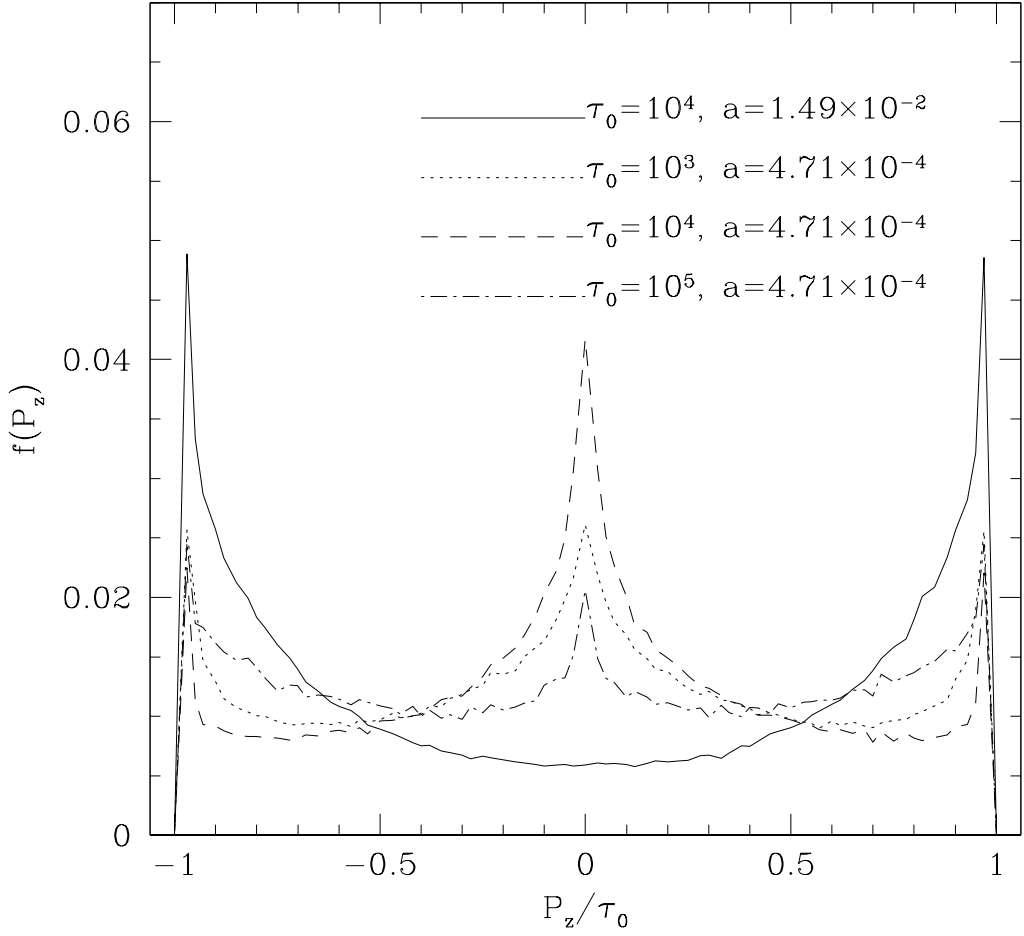}
\figcaption[f2.ps]{Last scattering positions of escaping photons.
We can see the two peaks at $P_z/\tau_0 \approx 0$
and $|P_z|/\tau_0 \approx 0.95$.
See the text for the detailed explanation for the origin of these peaks.
\label{f2}}
\end{figure}

\begin{figure}
\plotone{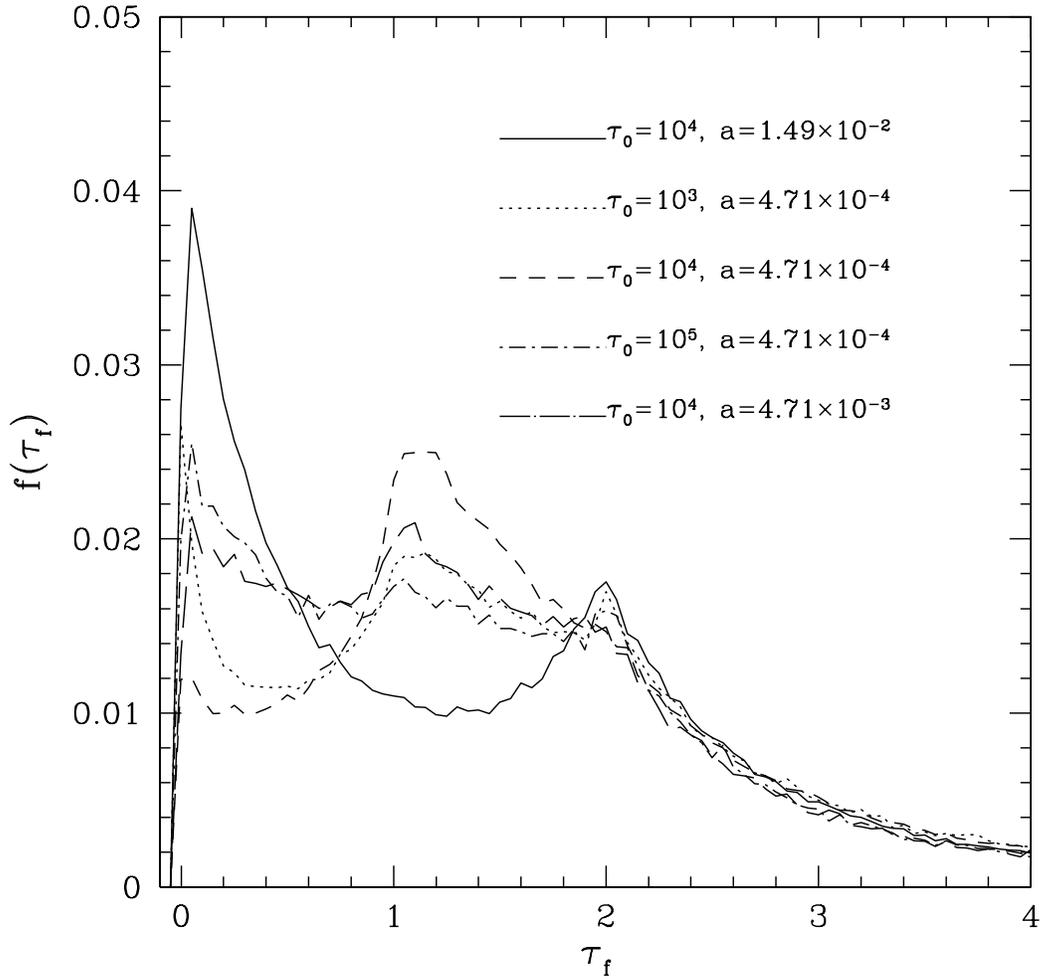}
\figcaption[f3.ps]{Distributions of photon path lengths just before
escape. The horizontal axis represents path lengths $\tau_f$,
which is defined by the transverse optical depth in units of $\tau_0$
for the escaping photons in the slab after their last scattering.
The vertical axis represents the probability density
for a photon to have a last scattering path length in a range
$(\tau_f,\tau_f+d\tau_f$).
There are three kinds of peaks:
the first is at $\tau_f \approx 0$, the second is at $1<\tau_f<2$,
and the third is at $\tau_f \approx 2$.
\label{f3}}
\end{figure}

\begin{figure}
\plotone{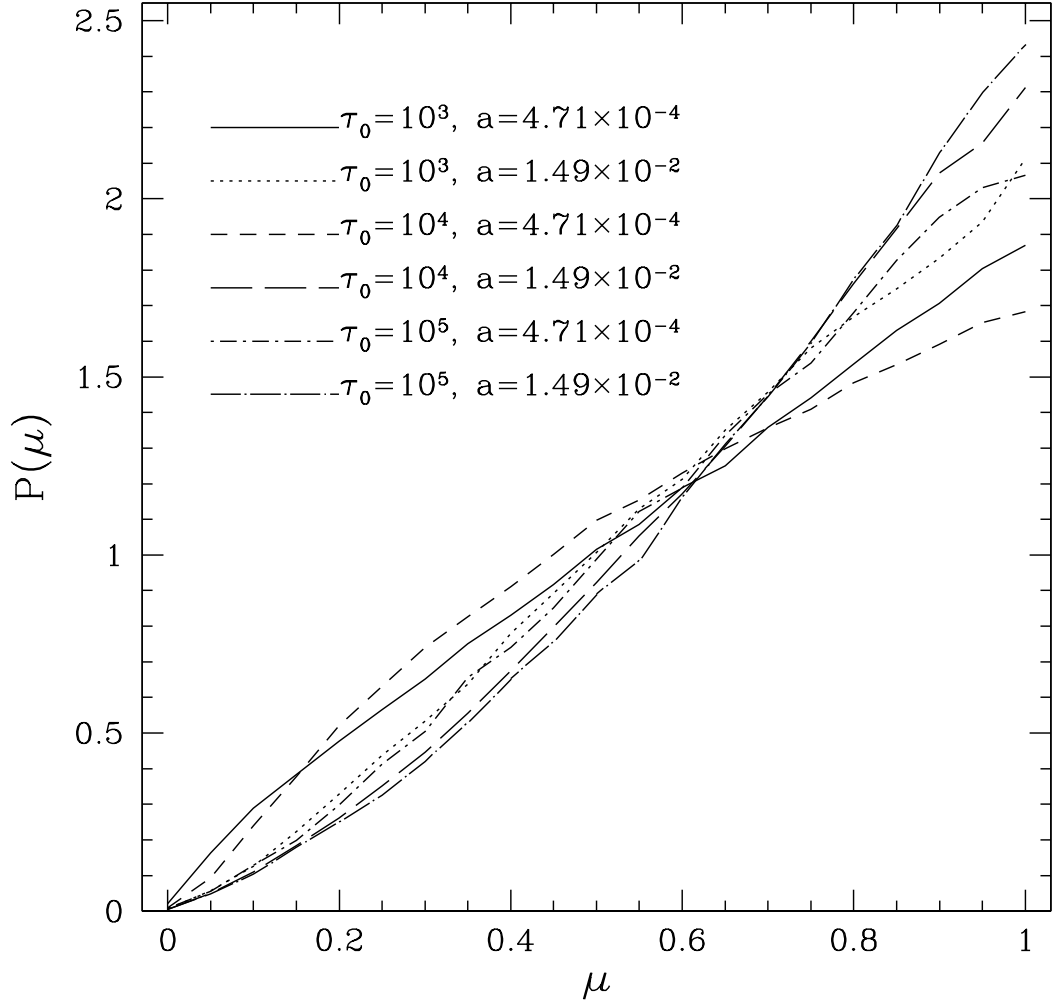}
\figcaption[f4.ps]{Angular distributions of emergent photons for various
optical depths $\tau_0$ and Voigt parameters $a$ of the scattering medium.
For isotropic distribution, $P(\mu )\propto \mu$.
We note that
the curves become convex downward when $a\tau_0$ is large,
whereas they become convex upward when $a\tau_0$ is smaller.
\label{f4}}
\end{figure}

\begin{figure}
\plotone{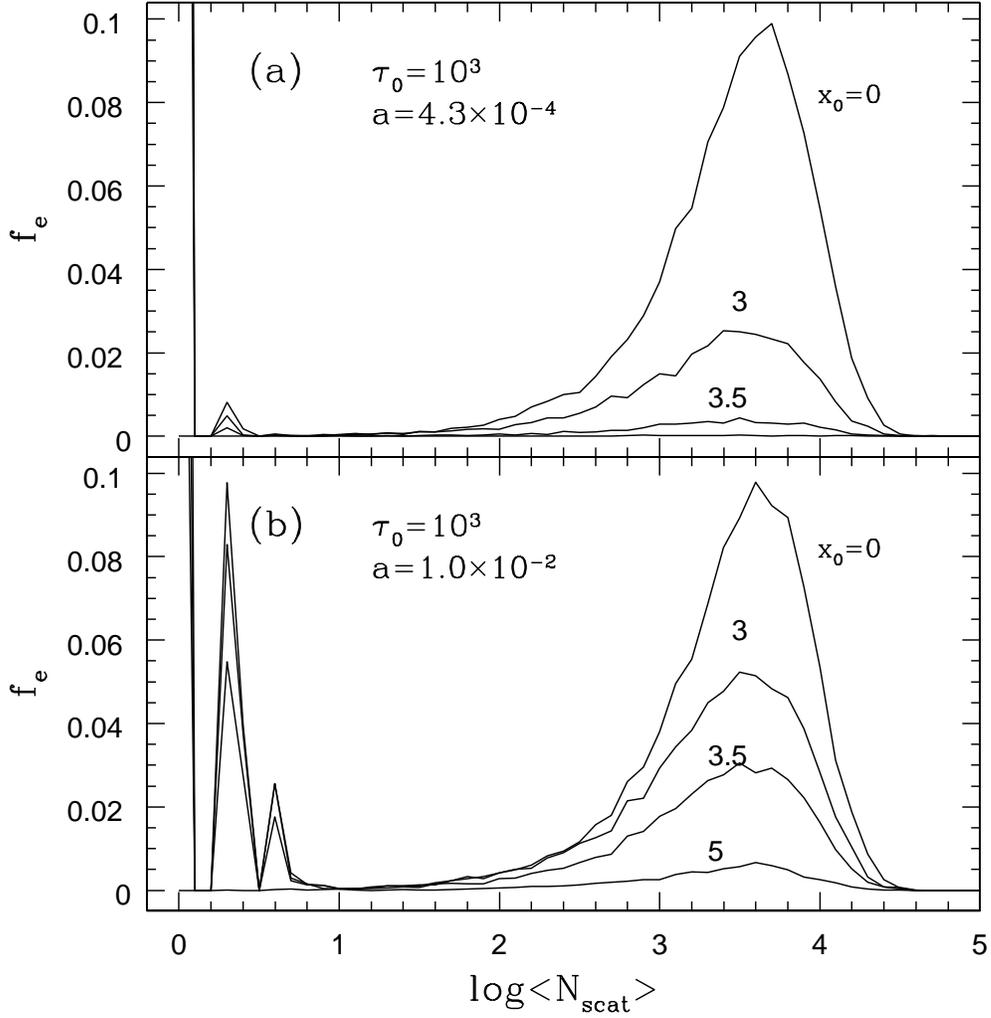}
\figcaption[f5.ps]{
Distributions of the mean number of scatterings for emergent Ly$\alpha$
photons. In the top panel (a), $\tau_0=10^3$ and $a=4.3\times10^{-4}$;
in the bottom panel (b), $\tau_0=10^3$ and $a=1.0\times10^{-2}$.
\label{f5}}
\end{figure}

\clearpage

\begin{figure}
\plotone{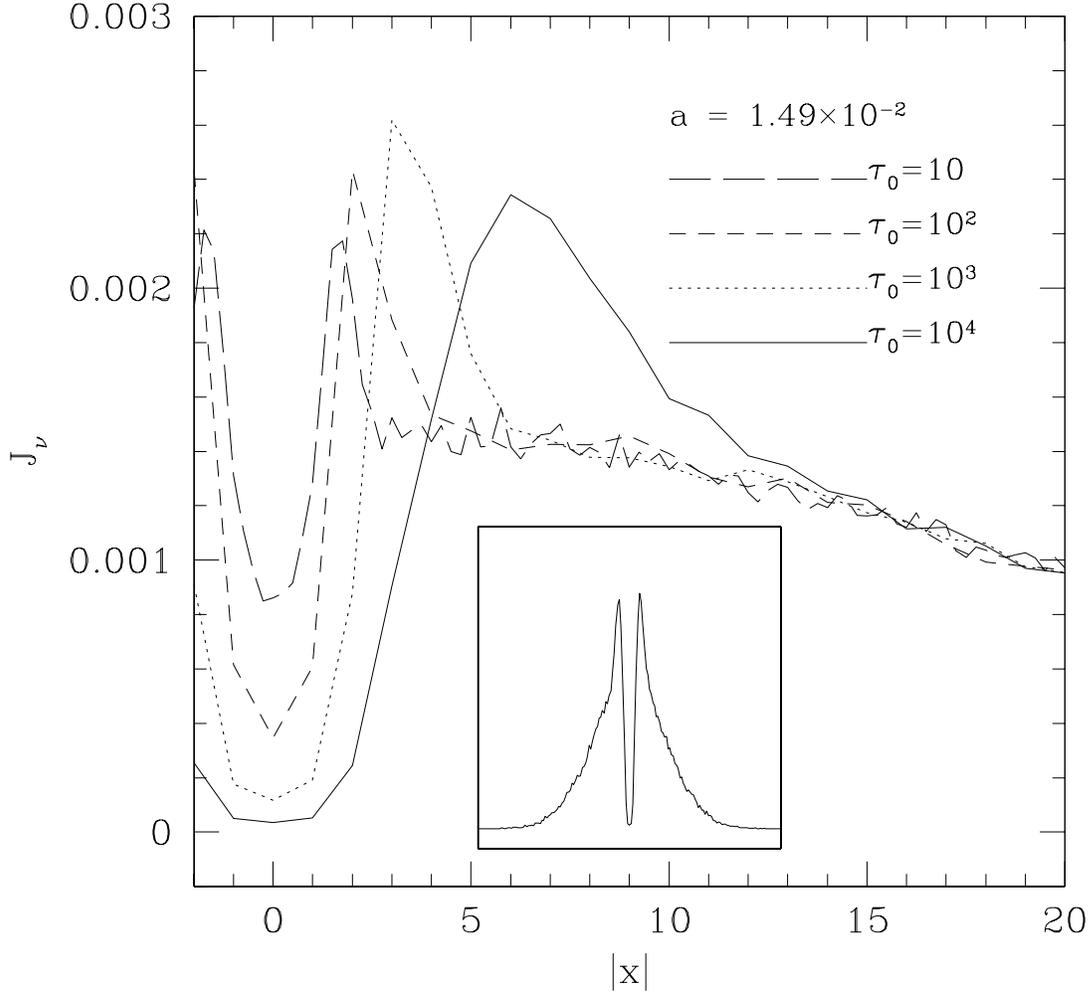}
\figcaption[f6.ps]{Emergent Ly$\alpha$ profiles for the cases of
a hot photon source surrounded by a cold medium,
whose line center optical depths are shown in the figure.
The Voigt parameter $a=1.49\times10^{-2}$ corresponds to $T=10\ {\rm K}$.
Hence $\Delta x=1$ corresponds to $\Delta\lambda = 0.0016{\rm\ \AA}$
in the rest frame.
The Ly$\alpha$ profile of the source is assumed to be a Gaussian,
$\exp(-x^2/x_s^2)$,
where $x_s=30$. We show the grobal profile in a small box in the figure.
The total flux is normalized to be $1/4\pi$.
The profiles are characterized by double peaks and the absorption trough 
around the center.
The absorption trough ends at $x\simeq \pm 3$, and hence the full
width $\Delta x\simeq 6$. The Ly$\alpha$ profile in the small box
is that for $\tau_0=10^4$, which shows a characteristic double peaks and
the absorption trough at the line center.
\label{f6}}
\end{figure}

\begin{figure}
\plotone{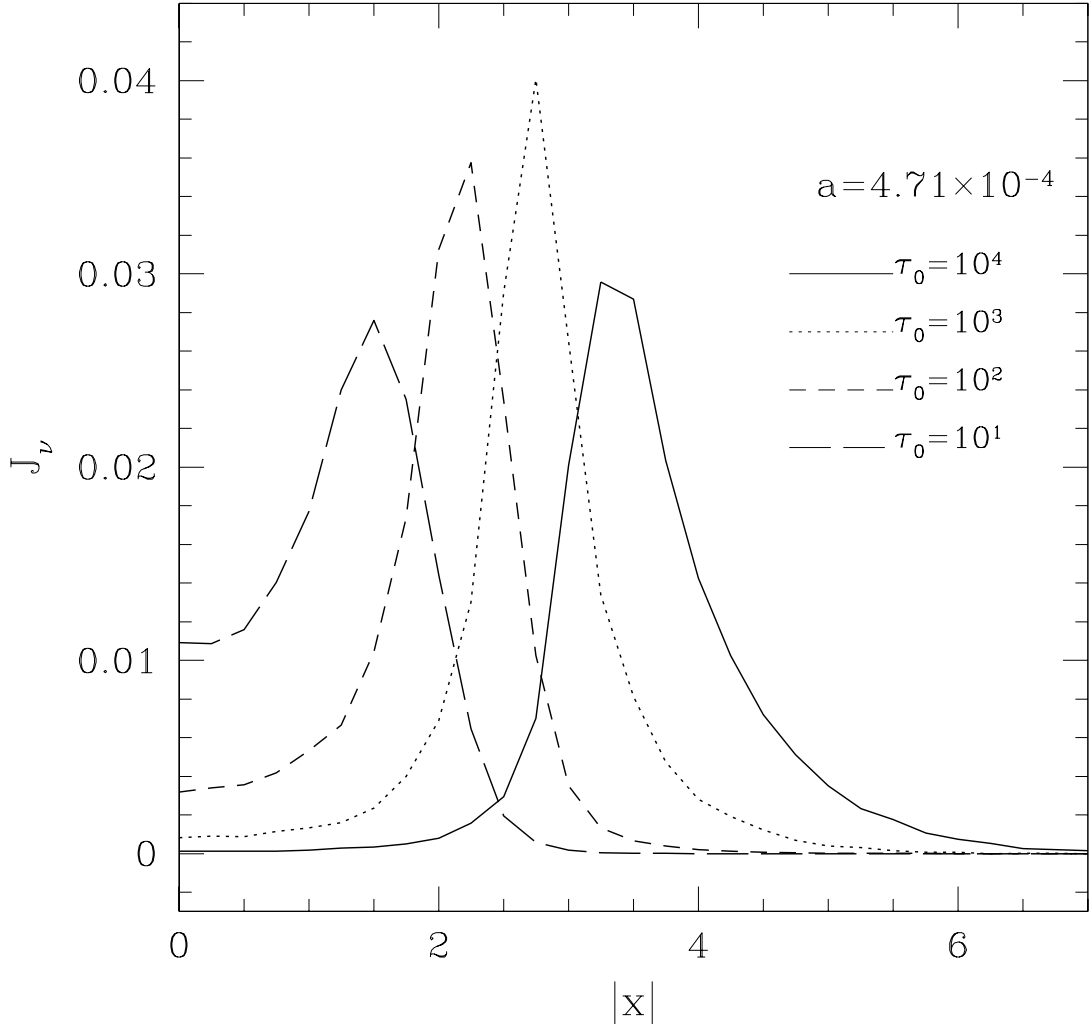}
\figcaption[f7.ps]{Emergent Ly$\alpha$ profiles for Case II,
where monochromatic ($x_0=0$) sources lie in the center and at and
are surrounded by a hot medium.
We assume that the slab-like scattering medium with the Voigt
parameter $a=1.49\times 10^{-2}$ or the temperature $T=10\ {\rm K}$
typical of \ion{H}{2} regions.
We note that the profiles are symmetric to the origin.
The central trough widens as the line center optical depth increases.
Here $\Delta x=1$ corresponds to $\Delta\lambda=0.05{\rm\ \AA}$
in the rest frame. \label{f7}}
\end{figure}

\begin{figure}
\plotone{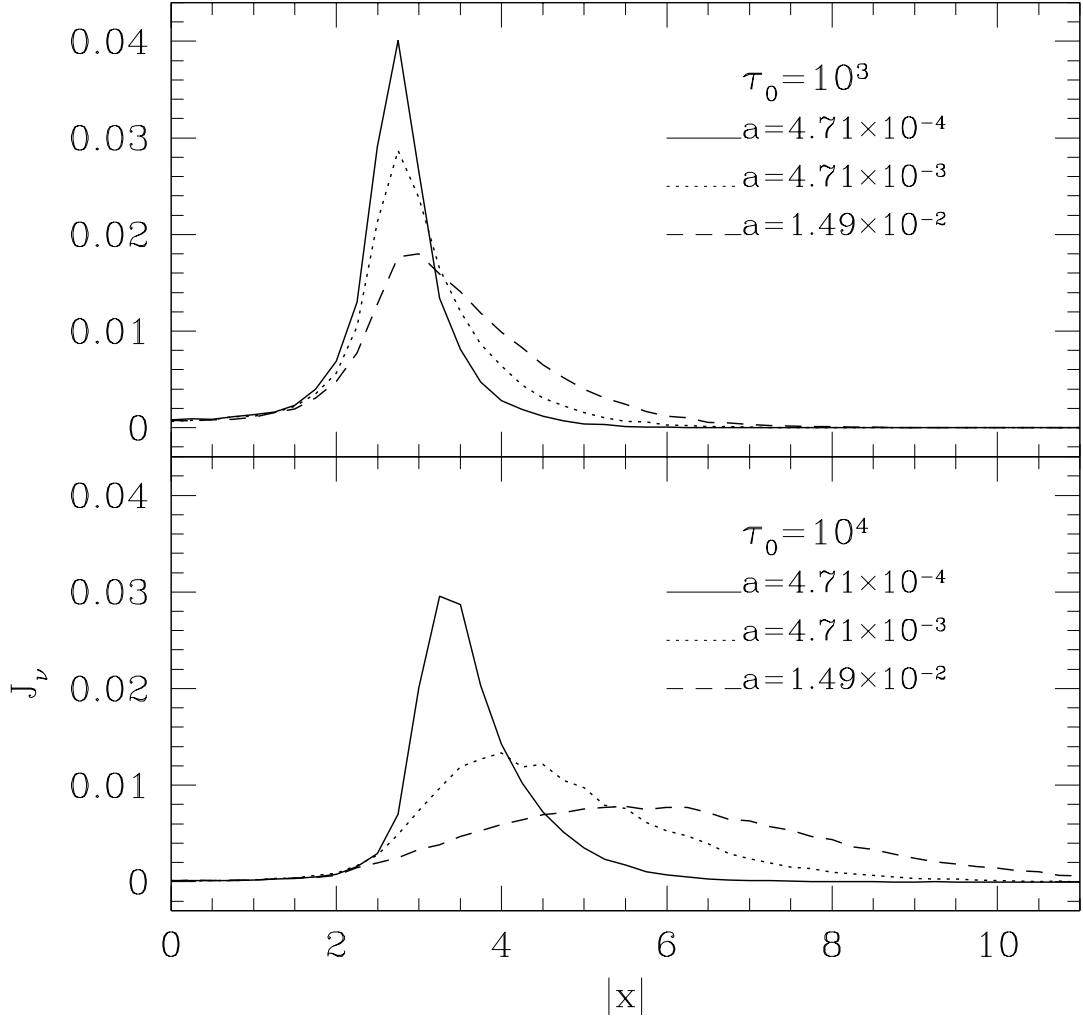}
\figcaption[f8.ps]{Emergent Ly$\alpha$ profiles for sets of
parameters of the scattering media. Here the monochromatic Ly$\alpha$
source is located at the center of the slab-like medium.
In the figure we denote the Voigt parameters and line center optical depths
of media by $a$ and $\tau_0$.
The profiles are symmetric to $x=0$ and
normalized so that the total flux becomes $1/4\pi$, and
symmetric to $x=0$. Note that when $a\tau_0$ is large, wing scattering
dominates the line transfer and results in broad bumps.
Here $\Delta x=1$ corresponds to $\Delta\lambda=0.05{\rm\ \AA}$ in
the rest frame. \label{f8}}
\end{figure}

\begin{figure}
\plotone{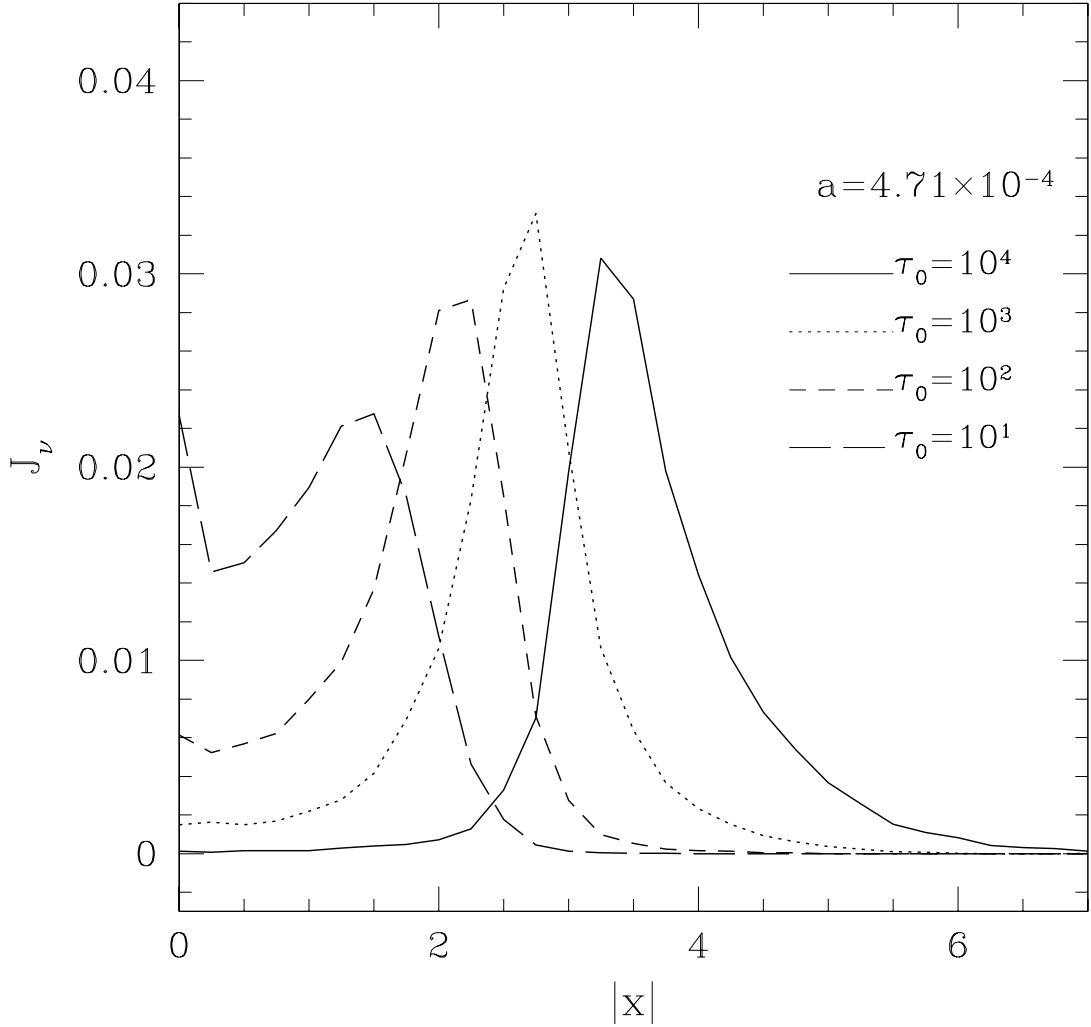}
\figcaption[f9.ps]{
Emergent Ly$\alpha$ profiles for Case III, in which monochromatic
Ly$\alpha$ sources are uniformly distributed in a hot slab-like medium.
We assume here the Voigt parameter of the medium is $a=4.71\times10^{-4}$
which corresponds to $T=10^4{\rm\ K}$, which is a typical value
for \ion{H}{2} regions.
The profiles are symmetric to the origin.
Comparing with the midplane case, we see that peaks become
fatter to the line center. Note that here $\Delta x=1$ corresponds
to $\Delta\lambda=0.05{\rm\ \AA}$ in the rest frame. \label{f9}}
\end{figure}

\begin{figure}
\plotone{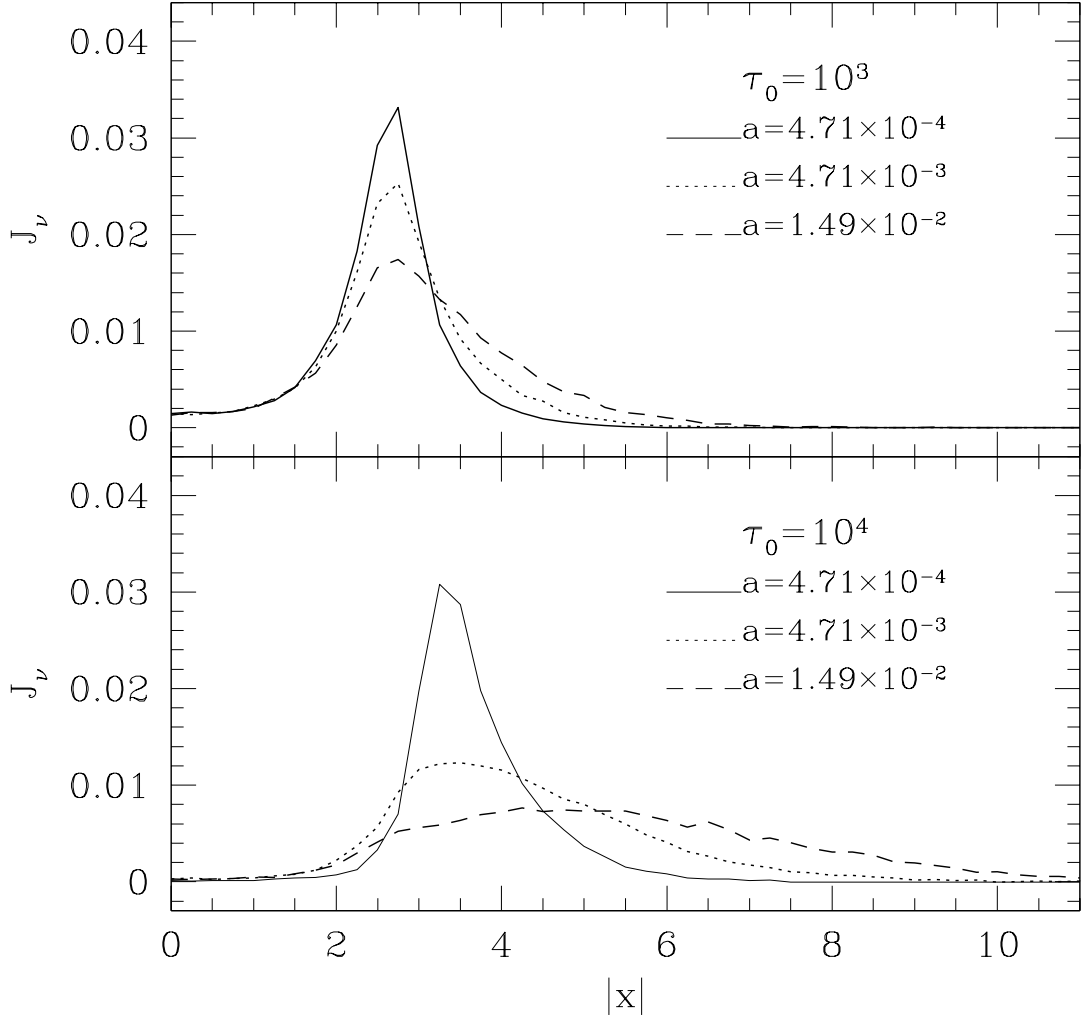}
\figcaption[f10.ps]{
Emergent profiles for various parameters of scattering media.
Here the monochromatic Ly$\alpha$ source is uniformly distributed
in the slab-like medium. We denote in the figure the Voigt parameters
and line center optical depths of media.
The profiles are normalized so that the total flux is $1/4\pi$,
and symmetric to $x=0$. Note that wing scatterings dominate
the line transfer and result in broad bumps.
The peaks are fatter to the line center that those
of Case II, the midplane case. Here $\Delta x=1$ corresponds
to $\Delta\lambda=0.05{\rm\ \AA}$ in the rest frame. \label{f10}}
\end{figure}


\begin{figure}
\plotone{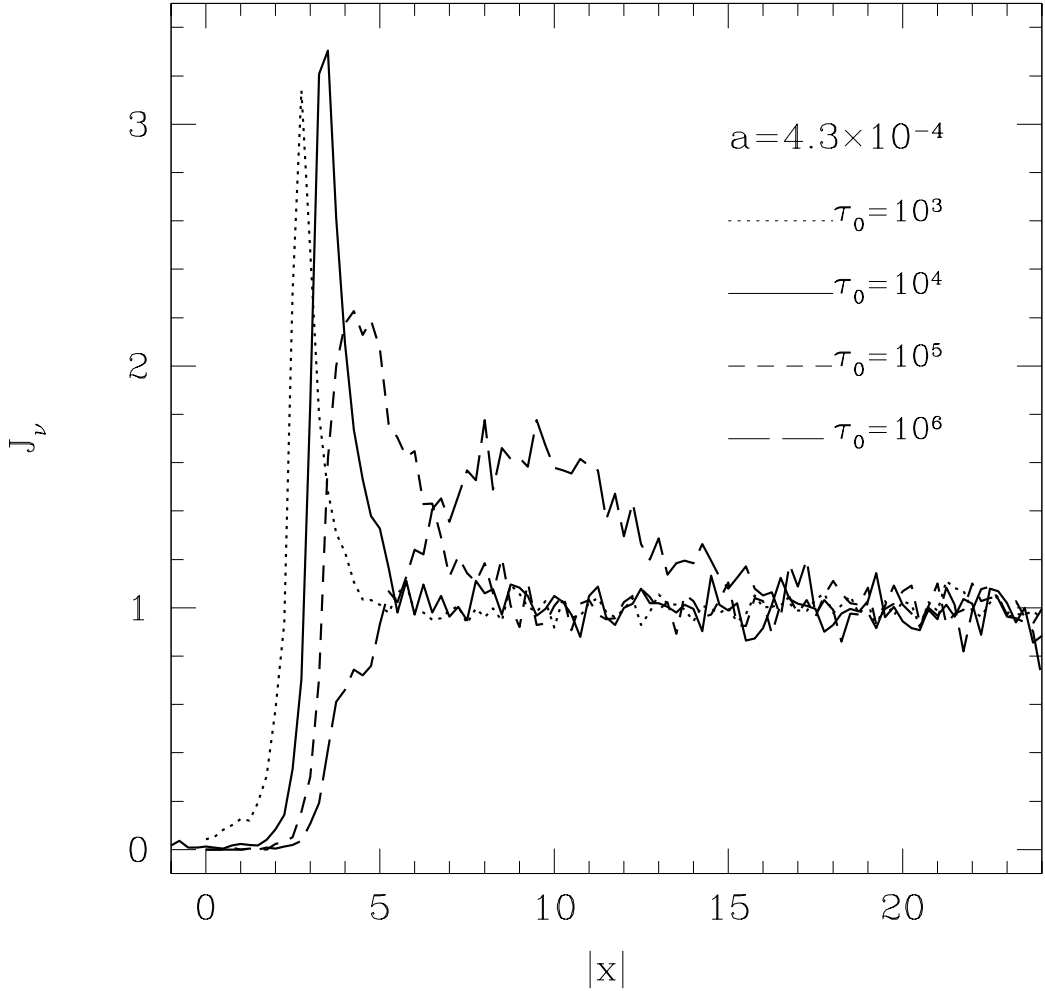}
\figcaption[f11.ps]{Emergent Ly$\alpha$ profiles for Case IV, in which
the continuum source is located at the center of the slab-like hot medium
with the Voigt parameter $a=4.3\times10^{-4}$. We show the effects of
the line center optical depths of the scattering media on the shapes
of the Ly$\alpha$ lines. The broader absorption troughs
appear as $\tau_0$ grows larger. Here $\Delta x=1$ corresponds
to $\Delta\lambda=0.05{\rm\ \AA}$ in the rest frame, and therefore
the width of the trough is about $0.3-0.5{\rm\ \AA}$.
\label{f11}}
\end{figure}

\begin{figure}
\plotone{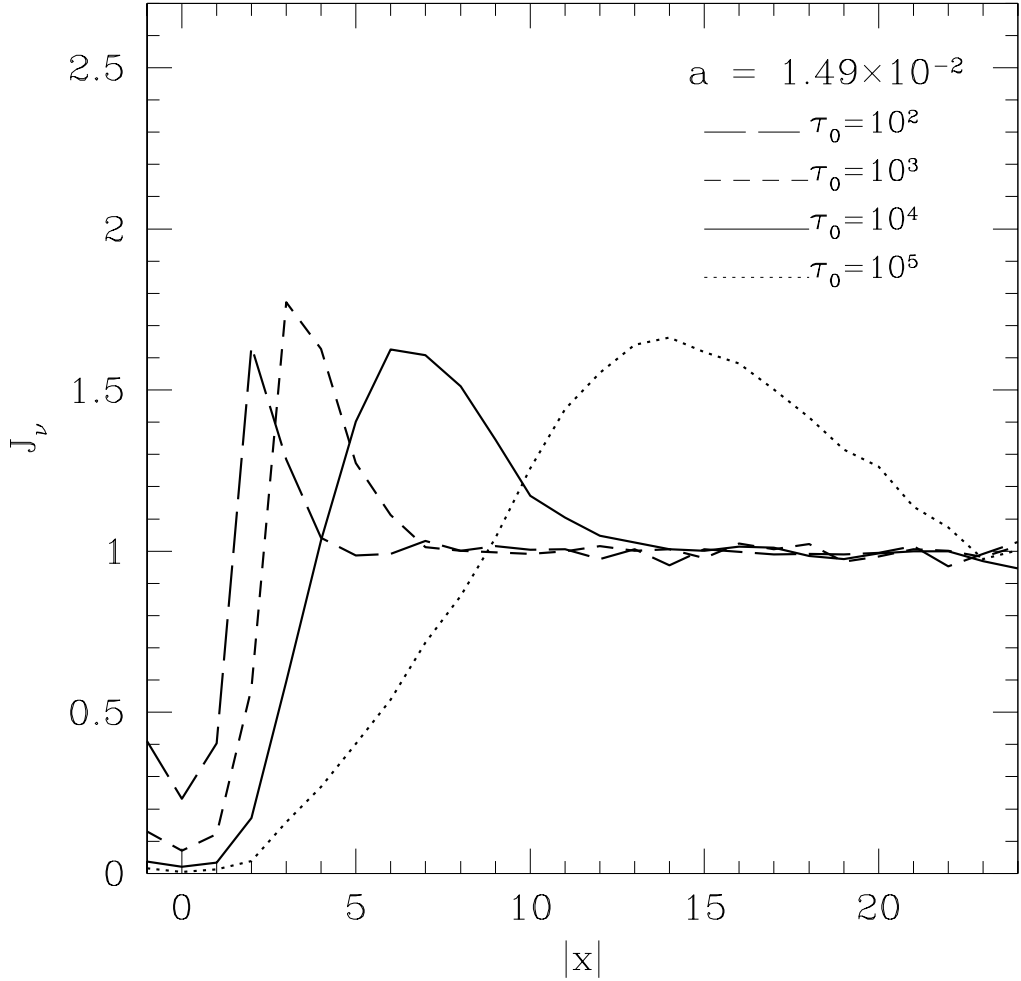}
\figcaption[f12.ps]{Emergent profiles for Case V, in which
the continuum source is located at the center of the slab-like cold medium
with temperature $T=10{\rm\ K}$. We show the effects of the line center
optical depths of the scattering media.
As $\tau_0$ grows large, there appear the absorption troughs which are
much broader than those in Case IV for the same $\tau_0$.
Here $\Delta x=1$ corresponds to $\Delta\lambda=0.0016{\rm\ \AA}$ in the rest frame,
and therefore the width of the trough ranges $0.01 - 0.03{\rm\ \AA}$.
\label{f12}}
\end{figure}

\clearpage

\begin{deluxetable}{cccccccc}
\tabletypesize{\scriptsize}
\tablecolumns{8}
\tablewidth{0pt}
\tablecaption{Ly$\alpha$ Emission Features of Starbursting Galaxies \label{tab1}
}
\tablehead{
\colhead{}    & \colhead{} & \multicolumn{2}{c}{scatterer} &  \colhead{} &
\multicolumn{2}{c}{trough} & \colhead{} \\
\cline{3-4} \cline{6-7} \\
\colhead{Stage} & \colhead{Case}  & \colhead{$\tau_0$}   & \colhead{$T$} &
\colhead{}  & \colhead{$\Delta\lambda_0 ({\rm \AA}) $}   & \colhead{$\Delta
\lambda_3 ({\rm \AA})$}  & \colhead{profile type\tablenotemark{4}}}
\startdata
(a) &  (I+V)\tablenotemark{3} & $10^{3-5}$ & 10 & & 0.01-0.03 & 0.04-0.13 & T+DP
\\
(b) & II+IV & $10^{3-5}$ & $10^4$ &      & 0.3-0.5  & 1.2-2.0 & T+DP\\
(c) & (II+IV)\tablenotemark{1} & \nodata    & \nodata& & \nodata &\nodata & ST+P
 Cyg\\
    & III\tablenotemark{2} & $10^{3-5}$ & $10^4$ & & 0.25 & 1 & T+DP\\
(d)\tablenotemark{5} & VI & $10^{6-9}$ & $10^{1-2}$ & & \nodata & \nodata & P Cy
g \\
(d) & I\tablenotemark{3}, V\tablenotemark{3} & $10^{3-5}$ & 10 & & 0.01-0.03  &
0.04-0.13 & T+DP\\
\hline

\multicolumn{2}{c}{Case I} & \multicolumn{6}{l}{Hot source at the center surroun
ded by cold scattering media}\\
\multicolumn{2}{c}{Case II} & \multicolumn{6}{l}{Monochromatic hot source at the
 center of a partially ionized hot medium}\\
\multicolumn{2}{c}{Case III} & \multicolumn{6}{l}{Distributed source in a slab w
ith $\tau_0<10^5$}\\
\multicolumn{2}{c}{Case IV} & \multicolumn{6}{l}{Continuum source at the center
of a hot scatterer}\\
\multicolumn{2}{c}{Case V} & \multicolumn{6}{l}{Continuum source at the center o
f a cold scatterer}\\
\multicolumn{2}{c}{Case VI} & \multicolumn{6}{l}{Scattered by expanding
supershell (Future Work)}\\
\enddata

\tablenotetext{1}{ Major emission line}
\tablenotetext{2}{ Minor emission line}
\tablenotetext{3}{ For reality $\tau_0=10^{6-9}$ should be achieved.}
\tablenotetext{4}{ T = Absorption trough; DP = Double peaks; P Cyg = P Cyg type
profile}
\tablenotetext{5}{ Future Work}
\end{deluxetable}


\begin{deluxetable}{cccccccr}
\tabletypesize{\scriptsize}
\tablecolumns{8}
\tablewidth{0pt}
\tablecaption{Spectral Capabilities of Devices \label{table2}}
\tablehead{
 & & Instrument & Spectrograph & & Range  & Resolution & \\
 & &            &              & & (${\rm \AA}$)& (${\rm \AA}$)   &  }
\startdata
 & & FUSE\tablenotemark{1} & HIST & & 905-1187  & 0.05-0.06 & \\
 & & HST/STIS\tablenotemark{2} & G140H & & 1140-1740 & 0.007-0.01 & \\
 & & Keck\tablenotemark{3} & LRIS & & 3900-11000 & 2-5 & \\
\enddata
\tablenotetext{1}{The Fuse Observer's Guide Version 2.1}
\tablenotetext{2}{STIS Instrument Handbook v4.1}
\tablenotetext{3}{\citet{oke95}}

\end{deluxetable}

\end{document}